# Large Language Models in Healthcare

**Mohammed Al-Garadi[1], Tushar Mungle[2], Abdulaziz Ahmed [5,6], Abeed Sarker [3,4] Zhuqi Miao[7,] Michael E. Matheny [1,8,9,10]**

[1] Department of Biomedical Informatics, Vanderbilt University Medical Center, Nashville, TN, USA.
[2] Department of Medicine, Stanford University, Stanford, CA, USA.
[3] Department of Biomedical Informatics, Emory University School of Medicine, Atlanta, GA, USA.
[4] Department of Biomedical Engineering, Georgia Institute of Technology Atlanta, GA, USA.
[5] Department of Health Services Administration, UAB, Birmingham, AL, USA
[6] Department of Biomedical Informatics and Data Science, UAB, Birmingham, AL, USA
[7] School of Business, The State University of New York at New Paltz, New Paltz, NY, USA.
[8] Department of Medicine, Vanderbilt University Medical Center, Nashville, TN, USA.
[9] Department of Biostatistics, Vanderbilt University Medical Center, Nashville, TN, USA.
[10] Geriatric Research Education and Clinical Care Service, Tennessee Valley Healthcare System VA, Nashville, TN, USA.

**Abstract**
Large language models (LLMs) hold promise for transforming healthcare, from streamlining administrative and clinical workflows to enriching patient engagement and advancing clinical decision-making. However, their successful integration requires rigorous development, adaptation, and evaluation strategies tailored to clinical needs. In this Review, we highlight recent advancements, explore emerging opportunities for LLM-driven innovation, and propose a framework for their responsible implementation in healthcare settings. We examine strategies for adapting LLMs to domain-specific healthcare tasks, such as fine-tuning, prompt engineering, and multimodal integration with electronic health records. We also summarize various evaluation metrics tailored to healthcare, addressing clinical accuracy, fairness, robustness, and patient outcomes. Furthermore, we discuss the challenges associated with deploying LLMs in healthcare— including data privacy, bias mitigation, regulatory compliance, and computational sustainability—and underscore the need for interdisciplinary collaboration. Finally, these challenges present promising future research directions for advancing LLM implementation in clinical settings and healthcare.

## 1. Introduction

Recent advances in large language models (LLMs) have shown powerful generative capabilities, drawing widespread attention worldwide.[1] The generative capabilities of these models, which enable the production of novel, coherent text, highlight the technology's potential to augment human thought and work across various industries, thereby driving progress in productivity, creativity, and innovation.[2] The healthcare sector represents a promising field for applying generative LLMs to improve workflows and outcomes.[3,4] The most immediately feasible application appears to be utilizing these models to automate routine administrative tasks, thus relieving clinical staff of clerical burdens and enabling them to focus their medical expertise on patients[5,6]. By automating documentation, scheduling, and billing tasks, healthcare providers can improve efficiency, reduce errors, and allocate more time to valuable clinical activities.[7]

Beyond administrative applications, LLMs have the potential to transform healthcare by leveraging their advanced natural language understanding to process medical information, support decision-making, advance research innovation, and enhance communication.[3,8,9] These models have demonstrated significant potential by combining knowledge from diverse sources to generate coherent insights, and improve the efficiency of healthcare systems[3,4,10]. Their capacity to adapt across various domains parallels advancements in genomic modeling, such as the recently developed genomic foundation models, which decode DNA's informational complexity to enable predictive and generative tasks in genomics.[11] These models demonstrate how domain-specific foundation models can harness the power of hybrid computational architectures to optimize efficiency and scalability. This paves the way for innovations in precision healthcare. For instance, a cardiology-focused model can swiftly identify at-risk patients and support earlier interventions. Meanwhile, LLMs can integrate vast medical knowledge to provide real-time decision support, research synthesis, and patient triage.[3,4,10]

However, the responsible adoption of generative LLMs in healthcare must couple technological enthusiasm with a pragmatic focus on infrastructural, validation, and privacy and security prerequisites. Maximizing healthcare LLM benefits hinges on tackling governance challenges (privacy, accountability, bias) and real-world barriers: seamless workflow integration, validation against shifting clinical standards, and compatibility with aging IT systems. Absent scalable infrastructure, rigorous real-world testing, and regulatory alignment, even advanced AI risks irrelevance—unable to sustainably boost diagnostic accuracy, operational efficiency, or patient outcomes.[12]

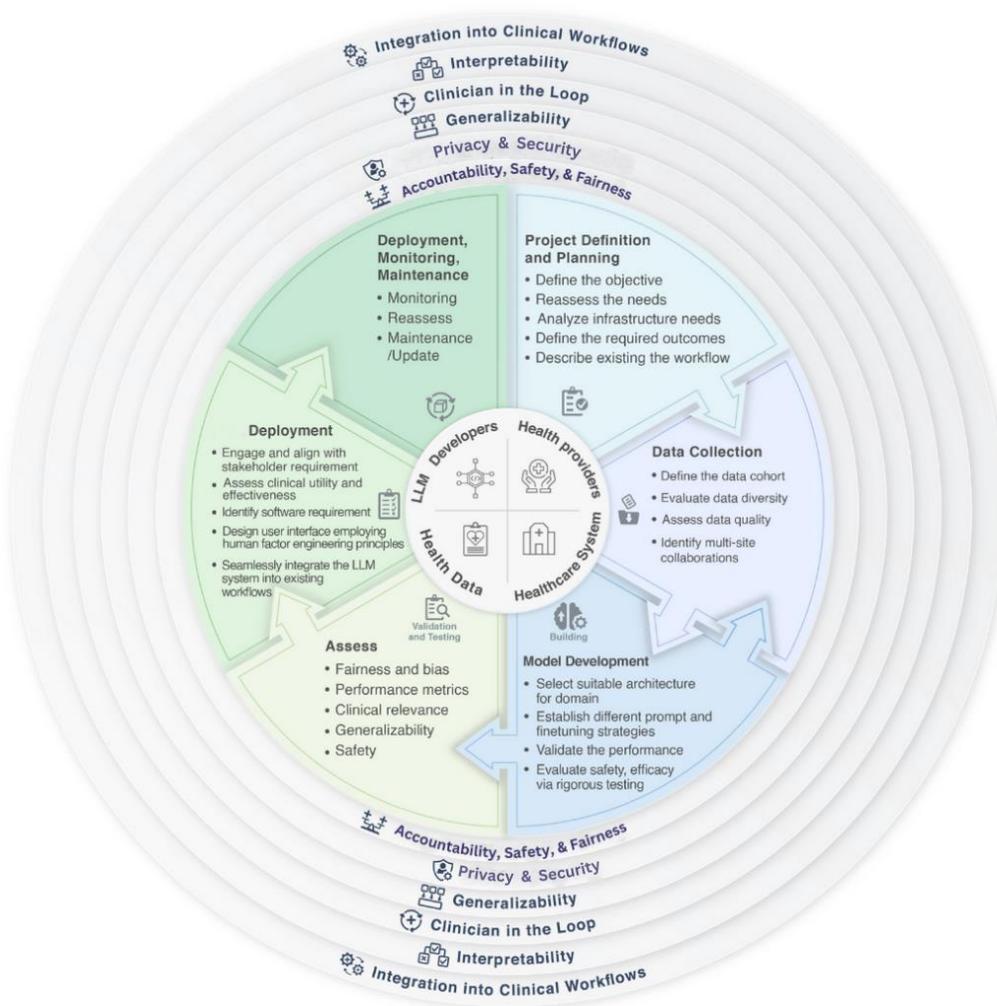

*Figure 1: A framework for integrating LLMs into healthcare. The inner core highlights collaboration between LLM developers, healthcare providers, and the healthcare system. The framework covers key stages: Planning, Data Collection, Model Development, Validation, Deployment, and Maintenance. The outer layers emphasize guiding principles, including Accountability, Privacy, Generalizability, Clinician Involvement, Interpretability, and Workflow Integration to ensure safe and effective implementation.*

## 2. LLM Integration into Healthcare Workflows

Recent LLMs promise to transform healthcare delivery, yet their integration demands thorough evaluation to maximize benefits and minimize risks. Figure 1 provides a framework for optimizing the responsible deployment of healthcare LLMs. Building upon prior models for implementing artificial intelligence in medicine, this framework promotes transparency, responsibility, confidentiality, accountability, and fairness across the AI lifecycle.[13-15] Therefore, the framework proposes an iterative paradigm for developing and implementing LLMs within healthcare analytics. Table 1 summarizes the lifecycle tasks and considerations for the implementation of healthcare LLMs, including five key phases: Planning, Development, Validation, Deployment, and Maintenance. Key considerations throughout the lifecycle include data privacy, regulatory compliance, accountability in decision-making, and alignment with clinical needs. The process involves curating high-quality data, constructing accurate and interpretable models, safeguarding patient privacy, and ensuring seamless integration into clinical workflows. Ensuring accountability at every stage requires transparent reporting, systematic auditing, and clear responsibility assignment for model predictions and clinical impact. Continuing iterative optimization, grounded in clinical data and user input, further enhances their reliability and robustness.

Healthcare LLMs require multidisciplinary collaboration to align technical performance with operational demands and clinical relevance—transforming validated systems into scalable tools that advance decision-support and measurable patient outcomes.

| Table 1: Essential Tasks at Each Phase of Lifecycle for Healthcare LLMs (Planning, Development, Validation, Deployment, and Maintenance), Associated with Figure 1. | |
|---|---|
| **Task** | **Specific Considerations for Healthcare LLMs** |
| 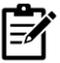 **Project Definition and Planning** | • **Defining Scope:** Identify targeted healthcare tasks for LLM relevance in clinical practice.<br>• **Clinical Needs Alignment:** Focus on clinician-defined tasks to ensure practical utility in healthcare applications.<br>• **Planning Bias Mitigation:** Create best practices to identify and mitigate biases in LLM outputs to ensure reasonable healthcare delivery.<br>• **Data Privacy Planning:** Establish governance and privacy measures to secure sensitive health data.<br>• **Regulatory Planning:** Analyze regulations and guidelines to enable ethical, legal LLM deployment in healthcare.<br>• **Setting Explainability Requirements:** Determine necessary explainability levels to foster trust.<br>• **Assessing Infrastructure and Planning Integration:** Evaluate IT systems to enable seamless LLM integration. |
| 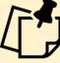 **Data Collection and Curation** | • **Data Representativeness:** Ensure broad demographic representation in training data.<br>• **Data Quality:** Curate clinically relevant, accurate data.<br>• **Data Breadth:** Include health conditions for comprehensive model understanding.<br>• **Clinician-Aligned Data Collection:** Involve clinicians to ensure contextual accuracy.<br>• **Workflow-Aware Data Collection:** Gather data aligned with existing clinical workflows.<br>• **Data Privacy**: Strict adherence to privacy norms for sensitive health information.<br>• **Data Governance**: Enforce robust data governance frameworks for compliance and security |
| 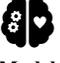 **Model Development and Validation** | • **Patient-Centric Design:** Incorporate patient feedback during development to ensure LLMs prioritize patient outcomes and preferences.<br>• **Clinician In-loop Design:** Continuously involve clinicians to align with practical clinical needs.<br>• **Performance Evaluation:** Implement comprehensive validation tests to assess LLM accuracy and reliability.<br>• **Clinical Utility Evaluation:** Assess and demonstrate the practical value of LLMs in enhancing clinical decision-making and patient care.<br>• **Interpretability:** Develop interpretable LLMs to enable clinician understanding and trust.<br>• **Scalability:** Design flexible, scalable LLMs capable of handling increasing data volume and complexity.<br>• **Uncertainty Quantification:** Integrate methods to quantify prediction uncertainty and provide risk assessments.<br>• **Hallucination Control:** Minimize false or misleading output generation to ensure reliability.<br>• **Regulatory Compliance:** Rigorously test to prevent data leaks and ensure compliance with regulations.<br>• **Continuous Learning**: Enable ongoing learning from new data to improve performance.<br>• **Integration and Deployment:** Focus on easy integration into healthcare IT systems and workflows. |
| 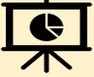 **Deployment** | - **Patient Outcomes:** Monitor changes in patient outcomes that may indicate effectiveness in improving care.<br>- **User Experience:** Gather clinician feedback on usability, ease of integration, and overall satisfaction through surveys and interviews.<br>- **Clinical Decision Support**: Measure the LLM's impact on diagnosis, treatment recommendations, and patient management through feasibility studies.<br>- **Privacy**: Pilot deployments should adhere strictly to ethical and privacy standards.<br>- **Clinical Interpretability:** The model's interpretations should be assessed to ensure they align with clinicians' understanding and expectations.<br>- **Workflow Integration:** The integration of the LLMs must be rigorously evaluated during the pilot testing period, with a critical analysis of its impact on clinical workflow efficiency, clinician usability, and the degree to which it supports or delays existing processes.<br>- **Outcomes and Metrics Focus:** During the pilot deployment, it's crucial to establish clear outcomes and metrics to measure the LLM's performance and its effect on patient care quality.<br>- **Continuous Feedback Loop:** There should be a system for continuous feedback from healthcare providers and patients during the pilot phase to refine and improve the LLM's performance.<br>- **Scalability Assessment:** The pilot phase should evaluate the LLM's scalability, ensuring it can handle increased data volume and user load without service or performance degradation. |
| 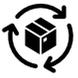 **Monitoring, Maintenance** | - **Clinical Collaboration:** Foster ongoing engagement with clinicians to ensure the LLM meets evolving clinical needs and benefits from frontline feedback, driving iterative improvements.<br>- **Health Outcome Tracking:** Commit to the continuous assessment of the LLM's impact on patient health outcomes, prioritizing improvements in patient care and treatment efficacy.<br>- **Integration Monitoring:** Continuously monitor the LLM's integration into healthcare workflows, ensuring seamless operation, identifying areas for improvement, and minimizing disruptions.<br>- **Feedback Loop:** Establish feedback mechanisms for healthcare professionals and patients to report issues or suggest enhancements, informing continuous model refinement.<br>- **Maintaining Interpretability**: As the model evolves, ensure it remains interpretable to clinicians, maintaining their trust and facilitating their understanding of model outputs.<br>- **Privacy Compliance:** Uphold ethical standards and privacy protections, adapting to evolving regulations and societal expectations to maintain patient trust and legal compliance.<br>- **Performance Monitoring:** Conduct ongoing evaluation of the LLM's performance across different populations and settings to ensure its effectiveness and adaptability.<br>- **Scalable Infrastructure Development:** Develop scalable infrastructure that supports efficient model updates and deployment across various healthcare settings, with a focus on modularity and automation.<br>- **Resilient Backup Systems:** Implement real-time data replication and disaster recovery strategies across distributed servers to ensure system reliability and data integrity.<br>- **Long-Term Impact Analysis:** Engage in long-term studies to gauge the broader impacts of the LLM on healthcare systems, patient outcomes, and healthcare delivery, facilitating strategic adjustments. |

The lifecycle framework for healthcare LLMs ensures systematic integration across all phases—from planning and data collection through model development, deployment, and maintenance. During the initial planning phase, institutions must establish robust governance protocols address data privacy and regulatory compliance. This phase lays the foundation by identifying clinical objectives that drive all subsequent activities, ensuring that LLM deployment is contextually relevant, clinically appropriate, and designed to minimize unintended adverse effects. In addition, institutions should evaluate technical factors such as infrastructure scalability, system interoperability with existing health information systems, and continuous monitoring mechanisms[1,3,4,10].

A critical component of the LLM integration process is the collaboration between clinicians and data scientists to curate representative datasets. This partnership facilitates the assembly of data that accurately mirrors real-world data and clinical workflows, thereby enhancing the model's relevance. In the model development phase, LLMs are rigorously fine-tuned to optimize clinical accuracy and interpretability by leveraging domain-specific data and robust explainability techniques. This process not only refines the model's understanding of medical terminology and context but also integrates additional factors such as data quality, privacy safeguards, regulatory compliance, and scalability. Advanced training protocols—often incorporating reinforcement learning from human feedback (RLHF)—help ensure that the model remains clinically relevant and that potential side effects, such as misinterpretation or unintended recommendations, are minimized. Pilot deployments in controlled clinical environments then allow for iterative refinement. Feedback from healthcare professionals is systematically analyzed alongside performance metrics, enabling further adjustments to enhance operational reliability and alignment with real-world clinical decision-making[3,4,10].

Finally, continuous monitoring and maintenance are essential to ensure that the deployed LLM adapts to evolving clinical requirements and regulatory landscapes. Through systematic performance evaluations and real-time feedback loops, healthcare systems can sustain model efficiency and integrity over time. This ongoing process is critical to achieving a long-term, effective integration of LLMs into clinical workflows, ultimately supporting improved patient care and healthcare outcomes[3,4,10].

## 3. LLM Functionalities for Healthcare Applications: Emerging Capabilities, Domain Adaptation, and Evaluation

Recent advances in LLMs have enabled innovative functionalities specifically tailored to healthcare needs. These emerging capabilities span across clinical, operational, and educational domains, offering novel solutions to longstanding healthcare challenges.

LLMs have the potential to transform healthcare through multiple clinical applications. In clinical documentation, LLMs integrated with automatic speech recognition can reduce the documentation burden on healthcare providers, who currently spend nearly 6 hours of their 11-hour workday on EHR documentation.[16] Training LLMs on unstructured clinical notes can enable a wide range of clinical and operational predictive tasks, making them effective all-purpose engines for real-time decision support within health systems.[17] Their capabilities extend to clinical text classification [18-20] and information extraction [21,22] across various disease domains, improving patient care through better data organization and analysis. In medical education [10,23,24], LLMs facilitate curriculum development and personalized learning, while also streamlining clinical trial recruitment by matching trial criteria with patient attributes [25-27]. LLM systems can enhance patient-provider communication by helping generate clear medical explanations and providing consistent messaging support, serving as a valuable tool to complement physician care in healthcare settings.[28-30]

The transformation continues into more specific areas of healthcare delivery and research. The GenePT model, which utilizes GPT-3.5 to create effective single-cell biology embeddings from literature, demonstrates LLMs' ability to derive biological insights without extensive experimental datasets.[31] In clinical settings, these models enhance medical coding and data capture by efficiently extracting key medical concepts from clinical notes, while enabling personalized care through patient-specific recommendations. [32] Within pharmaceutical research, LLMs accelerate drug discovery through comprehensive analysis of biomedical literature and chemical databases [33,34], and advance protein design and engineering by generating novel sequences beyond natural occurrences [35-39]. As virtual medical assistants [40,41], these systems show promise in providing patient education and medical triage, though their implementation requires careful consideration of accuracy, accountability, and privacy concerns. While these applications demonstrate LLMs' transformative potential in healthcare, their successful integration requires a systematic approach combining rigorous validation, interdisciplinary collaboration, and strict adherence to ethical guidelines—particularly in clinical settings where patient outcomes, and privacy protection are paramount.

### 3.1 Adapting LLMs for Domain-Specific Clinical Tasks

Leveraging pre-trained LLMs offers a transformative approach to developing healthcare-specific AI models. By adapting these models through fine-tuning, transfer learning, and prompt engineering, they can effectively address the distinct linguistic and operational requirements of healthcare. These adaptations enhance the precision and reliability of medical applications while balancing general language capabilities with domain-specific needs and adhering to privacy and regulatory standards.

Table 2 presents a summary of key strategies for adapting pre-trained LLMs to healthcare applications, organized across seven methodological categories: model foundation selection, domain adaptation techniques, knowledge transfer mechanisms, data-centric enhancements, multi-agent architectures, robustness optimization, and continuous adaptation protocols. It demonstrates how advanced techniques—from parameter-efficient fine-tuning and prompt engineering to knowledge distillation and differential privacy—address the complex requirements of clinical deployment. These strategies collectively enable the development of healthcare-specific LLM systems that combine clinical expertise with operational reliability. Through systematic implementation of these approaches, organizations will be able to create and validate LLM applications that meet the demands of healthcare environments.[7,42,43].

| \multicolumn{3}{c}{**Table 2 Approaches for Adapting General Pre-trained LLMs to the Specific Healthcare Domain**} | | |
|---|---|---|
| **Method Categories** | **Implementation Category** | **Approaches** |
| **1. Pre-trained Model Foundation** | *Base Model Selection* | An overall approach to LLM base model selection:<br>**Criteria**: Model size, training corpus relevance, and healthcare adaptability.<br>**Evaluate** existing LLMs tested in medical contexts.<br>**Analyze** capabilities and limitations, with examples of successful healthcare model applications to guide selection. |
| **2. Domain Adaptation Techniques** | *Prompt Engineering & Control* | **Prompt Design Strategies**<br>• Create clinical templates to standardize input formats.<br>• Engineer task-specific instructions to direct model outputs.<br>• Integrate safety constraints and robust safety mechanisms, including post-processing filters and human oversight, to prevent harmful recommendations.<br>**Chain-of-Thought Reasoning**<br>• Develop diagnostic prompts that guide the model through clinical reasoning steps.<br>Construct treatment logic chains to simulate decision-making processes. |
| | *In-Context Learning* | **Zero-Shot learning**<br>• Structure precise medical task descriptions with clear clinical parameters and expected outcomes to enable performance without prior training<br>• Apply established medical knowledge and reasoning patterns across different clinical scenarios to handle unfamiliar tasks effectively<br>**Few-Shot Learning**<br>• Provide limited medical examples to enable the model to perform specific tasks without extensive retraining.<br>• Leverage few-shot learning to generalize from limited data in medical tasks with scarce examples. |
| | *Fine-Tuning Approaches* | **Model Fine-Tuning (Parameter-Efficient Adaptation)**<br>• Apply fine-tuning methods such as LoRA (Low-Rank Adaptation) or quantized LoRA (QloRA) for efficient model updates.<br>• Use prompt-tuning to adjust model behavior with minimal parameter changes. |
| | *Interactive Enhancement* | **Reinforcement Learning**<br>• Incorporate expert feedback cautiously to refine model outputs, considering ethical implications and challenges in defining appropriate reward functions.<br>• Align model iterations with clinical guidelines and best practices, backed by multiple clinical validations<br>• Employ strictly controlled RL environments with robust safety protocols and maintain audit trails.<br>**Dialogue-Based Learning**<br>• Refine medical Q&A capabilities for precise information retrieval<br>• Adapt clinical conversations to align with practitioner communication styles. |
| | *Capability Expansion* | **Medical Knowledge Integration**<br>• Build dynamic knowledge bases to provide up-to-date information.<br>• Develop cross-referencing using techniques such as Retrieval-Augmented Generation (RAG) or Fusion-RAG (Enhanced RAG that combines multiple knowledge sources for better retrieval)<br>**Multimodal Adaptation**<br>• Explore integrating multimodal data to enhance diagnostic support, acknowledging the complexities of medical imaging interpretation. |
| **3. Knowledge Transfer & Enhancement** | *Knowledge Integration* | • **LLM-Generated Embeddings:** Utilize LLMs to create textual embeddings that enhance predictive accuracy and data analysis across various machine learning models.<br>• **Structured Data Integration**: Incorporate medical databases for extensive, context-rich information.<br>• **Cross-Domain Knowledge Preservation**: Maintain versatility by ensuring the retention and transfer of knowledge across different healthcare domains. |
| **4. Data-Centric Enhancement** | *Data Enrichment & Knowledge Base Integration* | **Medical Data Augmentation**<br>• Augment datasets to improve model training on medical tasks.<br>• Generate synthetic clinical data to simulate various scenarios, ensuring representation.<br>**Rare Case Data Simulation**<br>• Utilize data simulations of uncommon conditions to prepare the model for rare cases.<br>• Enhance the model's ability to recognize and respond to atypical presentations.<br>**Integration of External Knowledge Bases**<br>• Incorporate external medical knowledge bases for enriched information and context.<br>• Ensure the accuracy and reliability of integrated knowledge sources. |
| **5. Multi-Agent LLMs** | *Agent Architecture & Collaboration* | • **Workflow Analysis**: Explore modeling aspects of healthcare team workflows while identifying current limitations in automation and coordination.<br>• **Role-Specific Agent Development**: Design and refine AI-driven agents tailored for personalized assistance across different roles, such as case managers, physicians, and nurses.<br>• **Collaborative Process Simulation:** Model and simulate team-based workflows and care pathways to better capture the complexities of real-world clinical interactions.<br>• **Multi-Step Decision Support**: Facilitate structured decision-making with contextual guidance, recognizing that multi-agent LLM systems remain experimental and require rigorous validation |
| **6. Model Robustness and Efficiency Optimization** | *Resilience Enhancement* | **Adversarial Training**<br>• Implement medical-specific techniques to handle challenging inputs.<br>• Optimize error detection to minimize inaccuracies.<br>• Handle clinical edge cases effectively.<br>• Quantify uncertainty to assess confidence levels.<br>**Privacy Preservation**<br>• Integrate differential privacy to protect patient data.<br>• Apply anonymization techniques to safeguard personal health information (PHI).<br>• Ensure regulatory compliance with healthcare standards. |
| | *Efficiency Optimization* | **Knowledge Distillation**<br>• Compress medical knowledge selectively for efficient model size.<br>• Prune models for specific tasks to enhance performance.<br>• Optimize architecture for streamlined processing.<br>• Design efficient inference mechanisms to reduce latency.<br>**Performance Balancing**<br>• Analyze size versus accuracy to find optimal configurations.<br>• Optimize resources for cost-effective deployment.<br>• Reduce latency to improve user experience. |
| **7. Continuous Adaptation Framework** | *Adaptation Pipeline Management* | • Implement knowledge updates to keep information current.<br>• Monitor performance to identify areas for improvement.<br>• Iterate improvements based on feedback and new data.<br>• Incorporate expert feedback for ongoing refinement.<br>• Version control tracking: Both model iterations and training data versions. |

## 3.2 Evaluation of LLM Applications in Healthcare

In healthcare, traditional metrics such as accuracy and precision, while necessary, fall short of capturing the full spectrum of LLMs' reliability and ethical implications. A comprehensive evaluation framework is critical to address the unique demands of healthcare applications. Table 3 highlights focused metrics tailored for healthcare LLMs, detailing their objectives and practical applications[42,44-47]. Further, integrating a robust framework for prospective randomized controlled trials is vital. These trials should evaluate clinical outcomes with and without LLM integration, focusing on key areas, including efficiency gains, diagnostic and treatment accuracy, and patient satisfaction. Equally critical is the continuous monitoring of unintended consequences, particularly biases or disparities that may emerge from LLM deployment[48].

To ensure sustainable integration, post-deployment evaluations must systematically measure the long-term impacts of LLMs on healthcare delivery. These evaluations should measure improvements in patient outcomes, and clinician well-being. This approach aligns with the proposed milestones for responsible AI innovation, emphasizing bias mitigation, ethical adherence, and scalability tracking across various healthcare settings.[49]

*Table 3 Evaluation Metrics: Definitions and Uses[42,44-47]*

| Category | Metric | Definition | Uses |
|---|---|---|---|
| Basic Classification Metrics | Accuracy | The proportion of correctly predicted instances (true positives + true negatives) divided by the total number of predictions. | Evaluates overall correctness in tasks like classification and question answering. |
| | Precision (Positive Predictive Value) | The proportion of true positives among all positive predictions. | Indicates the accuracy of positive predictions. |
| | Recall (Sensitivity) | The proportion of true positives among all actual positives. | Measures how well the model identifies all relevant cases. |
| | F1-Score | The harmonic mean of precision and recall, balancing both metrics. | Measures classification performance, especially for imbalanced datasets. |
| | Confusion Matrix | A table displaying counts of true positives, true negatives, false positives, and false negatives. | Provides a comprehensive view of classification performance and allows calculation of metrics like precision, recall, and F1-score. |
| | Weighted Kappa | A statistical measure of inter-rater agreement that accounts for the degree of disagreement between raters, assigning weights to disagreements. | Commonly used in classification tasks involving ordered categories (e.g., disease severity ratings). Evaluates agreement between coders, annotators, or systems. |
| Text Generation Quality (Comparisons with Reference Outputs) | ROUGE (Recall-Oriented Understudy for Gisting Evaluation) | Compares overlapping n-grams, word sequences, and word pairs between the generated text and a reference. | Evaluates text summarization and machine translation quality. |
| | BLEU (Bilingual Evaluation Understudy) | Compares n-gram overlaps between predicted and reference translations, focusing on precision. Primarily used for machine translation evaluation. | Evaluates machine translation quality, assessing how closely the machine-generated text matches human translations. |
| | GLEU (Google Bilingual Evaluation Understudy) | A variation of BLEU that focuses on both precision and recall of n-grams, providing a more balanced measure for translation quality. | Used in machine translation evaluation, offering a more stable and fair assessment of quality, especially on sentences of varying length. |
| | Exact Match (EM) | A binary metric that evaluates whether the predicted output matches the reference answer exactly. | Common in question answering and text generation tasks. |
| Linguistic & Content Quality (Language Fluency, Comprehensibility, and Consistency) | Perplexity | The exponentiated average negative log-likelihood of the test set under the model. Lower perplexity indicates better language modeling ability. | Used in language modeling, lower perplexity indicates better generative models. |
| | Readability Scores | Quantitative measures (e.g., Flesch-Kincaid) indicating how easily an average reader can understand the generated text. Useful for patient-facing materials. | Evaluates simplified or educational content for patients or learners. |
| | Response Coherence | Evaluates the logical consistency and relevance between different parts of a response, ensuring it forms a cohesive and meaningful whole. | Applied in conversational AI to assess dialogue quality. Ensures responses are contextually and logically aligned with prior inputs. |
| | Consistency | Measures the degree to which a model produces stable and logically aligned outputs across similar or identical inputs. | Ensures reliability in decision-making tasks, reducing contradictory outputs. Evaluates stability in recommendations, classifications, or predictions. |

| Category | Metric | Description | Application |
|---|---|---|---|
| | *Linguistic Coverage* | The model process and generate various languages and dialects is crucial for healthcare applications | This capability enhances patient-provider communication, leading to improved health outcomes. |
| | *Redundancy Rate* | The proportion of repetitive or unnecessary content in the generated output. Lower redundancy indicates more concise and informative text. | Evaluates efficiency and relevance in text outputs. Assesses conciseness and informativeness of generated summaries or responses. |
| | Semantic Similarity Scores | Quantifies the semantic similarity between predicted outputs and reference texts using metrics like cosine similarity or STS scores. | Evaluates tasks like paraphrase detection, semantic matching, and medical coding accuracy (e.g., mapping to ICD/CPT codes). |
| *Clinical Decision & Risk Metrics* | *AUROC (Area Under the Receiver Operating Characteristic Curve)* | Measures the model's ability to distinguish between classes by plotting the true positive rate against the false positive rate across thresholds. Provides an aggregate measure of performance across all classification thresholds. | It is commonly used in diagnostic tasks and binary classification to assess discriminative ability. |
| | Concordance Index (C-Index) | A measure of the predictive accuracy of a risk score indicates the probability that, for a randomly selected pair of individuals, the one who experiences the event first had a higher predicted risk. | Commonly used in survival analysis to evaluate the discriminatory power of risk prediction models. |
| | *Calibration (ECE - Expected Calibration Error)* | Measures how well the predicted probabilities reflect actual outcomes by comparing predicted probabilities with observed frequencies. | Evaluates probabilistic models in tasks like risk prediction, ensuring confidence estimates are reliable. |
| | Brier Score | A metric that measures the mean squared difference between predicted probabilities and the actual binary outcomes. | Assesses the accuracy of probabilistic predictions, with lower scores indicating better calibration and sharper probability estimates. |
| | *Likelihood Ratios* | The ratio of the probability of a test result given the presence of a condition to the probability of the same result given its absence. Used in diagnostic models to interpret test results. | Used in diagnostic models to assess how much a test result will change the odds of having a condition. |
| | Triage Accuracy | Assesses how correctly the model assigns urgency or priority levels to patient cases, simulating triage decisions in a clinical setting. | Used in emergency medicine and risk stratification. |
| *Fairness & Biases Metrics* | *Demographic Parity Difference (DPD)* | Measures whether positive prediction rates are consistent across different demographic groups. Lower differences indicate more equitable predictions. | Ensures fairness in models by checking for equal chance of positive outcomes across groups, mitigating bias. |
| | Equalized Odds Difference (EOD) | Assesses whether the true positive and false positive rates are similar across demographic groups. Smaller differences indicate fairer performance. | Evaluates bias in classification models by ensuring error rates are equitable among groups. |
| *Efficiency & Usability Metrics* | *Compression Ratio* | The ratio of the length of the input text to the length of the output summary indicates how much the content has been condensed. | Evaluates text summarization efficiency by assessing how effectively information is condensed. |
| | Word Error Rate (WER) | The percentage of errors (substitutions, deletions, insertions) in a transcribed text compared to the reference text, calculated as (S + D + I)/N, where S = substitutions, D = deletions, I = insertions, N = number of words. | Assesses automatic speech recognition (ASR) systems. |
| | *Latency* | The time taken by a model to generate predictions or outputs. | Measures efficiency and responsiveness in real-time applications. |
| | Task Completion Rate | The proportion of tasks successfully completed by the system relative to the total tasks attempted, where success is defined by a predefined goal. | Evaluates effectiveness in task-based systems, such as virtual assistants or workflows. Indicates the model's ability to meet user objectives. |
| *Robustness & Safety Metrics* | *Adversarial Robustness Metrics* | Measures the model's performance degradation when exposed to adversarial inputs, such as attack success rate (percentage of successful attacks). | Assesses model robustness and resilience against adversarial attacks and input perturbations. |
| | Hallucination Rate | The frequency at which the model produces factually incorrect or fabricated information not supported by the source data or known medical facts. | Evaluates factuality and safety in generated content. |
| *Empathy and Trust Metrics* | *Empathy Scoring* | Measures the degree to which the model's responses demonstrate understanding, compassion, or emotional support—important for patient-provider communication. | Assesses conversational AI chatbot empathy. |
| | Trustworthiness Evaluations | Subjective or survey-based measures of how much clinicians or patients trust the model's outputs are critical for integrating AI into real-world healthcare workflows. | Important in tasks involving direct interaction with patients or providers to ensure the model is reliable and trusted. |

Table 3 outlines essential evaluation metrics and their applications of LLM for healthcare, emphasizing accuracy, fairness, and trust. Metrics like accuracy, precision, recall, and F1-score assess classification performance, while AUROC and calibration evaluate diagnostic and risk prediction reliability. Fairness measures, such as demographic parity difference and equalized odds difference, address biases across demographic groups, ensuring fair outcomes. Text-specific metrics like BLEU and ROUGE evaluate linguistic quality in summarization and translation, while robustness metrics gauge resilience to adversarial inputs.

Table 3 extends beyond technical performance to incorporate user-centric evaluation through efficiency metrics, empathy scoring, and trustworthiness assessments. Clinical risk metrics, including the concordance index and likelihood ratios, specifically address healthcare-specific requirements for longitudinal analysis and diagnostic reliability. Collectively, these various metrics enable the evaluation of healthcare LLMs, ensuring they meet the rigorous demands of clinical practice while maintaining patient-centered care and ethical standards.

| Table 4: Healthcare Task Categories with Metrics | | | |
|---|---|---|---|
| Healthcare Task Category | NLP/NLU Task Type | Quantitative Metrics | Qualitative Metrics |
| Clinical Knowledge Assessment | Medical Question Answering | Exact Match, F1 Score, ROUGE Scores, Calibration Metrics, Hallucination Rate, Demographic Parity Difference, Equalized Odds Difference | Expert evaluation on relevance, accuracy, and fairness; Trustworthiness Evaluations |
| Clinical Text Classification | Text Classification | Precision, Recall, F1-Score, AUROC, Calibration Metrics, Confusion Matrix, Adversarial Robustness Metrics, Demographic Parity Difference, Equalized Odds Difference | Clinician feedback on interpretability, clinical significance, and ethical considerations |
| Diagnostic Support | Differential Diagnosis Proposals | Diagnostic Accuracy, Sensitivity, Specificity, AUROC, Likelihood Ratios, Calibration Metrics, Hallucination Rate, Demographic Parity Difference, Equalized Odds Difference | Physician validation of suggestions; Safety and ethical evaluations |
| Symptom Analysis | Named Entity Recognition (NER) | Precision, Recall, F1-Score, Adversarial Robustness Metrics, Model Calibration under Noisy Inputs, Semantic Similarity Scores, Hallucination Rate, Demographic Parity Difference, Equalized Odds Difference | Expert assessment of appropriateness and fairness |
| Treatment Planning and Support | Care Plan Generation | ROUGE, BLEU Scores, Factual Accuracy, Consistency of Recommendations, Hallucination Rate, Readability Scores, Compression Ratio, Demographic Parity Difference, Equalized Odds Difference | Provider validation; Safety assessments; Trustworthiness Evaluations |
| Evidence Synthesis | Summarization | ROUGE Scores, Precision, Recall, F1-Score, Completeness of Synthesized Evidence, Hallucination Rate, Compression Ratio, Demographic Parity Difference, Equalized Odds Difference | Researcher review on comprehensiveness, relevance, and bias. |
| Administrative Automation | Clinical Documentation | Word Error Rate, Latency, Readability Scores, Hallucination Rate, Redundancy Rate, Demographic Parity Difference, Equalized Odds Difference | Workflow impact surveys; Trust Evaluations |
| Automated Medical Coding | Code Assignment | Exact Match Rate, Weighted Kappa, Precision, Recall, F1-Score, Semantic Similarity Scores, AUROC, Hierarchical Coding Accuracy | Coder feedback on usability, consistency, and clinical relevance |
| Patient Interaction | Conversational AI for Patient Questions | Task Completion Rate, Response Accuracy, Empathy Scoring, Hallucination Rate, Response Coherence, Demographic Parity Difference, Equalized Odds Difference | Patient feedback on trustworthiness and emotional engagement; User Satisfaction |
| Personalized Health Education | Content Simplification | Accuracy Scores, Readability Scores, Content Simplicity Metrics, Hallucination Rate | Feedback on clarity, usability, and safety; Learner feedback |

Table 4 links evaluation metrics to healthcare tasks, demonstrating their real-world applications. Quantitative metrics like F1-score, AUROC, and calibration assess performance across tasks, while fairness measures ensure ethical outcomes. Qualitative feedback, such as provider and patient evaluations, complements these metrics, enhancing trust and relevance in AI-driven healthcare solutions.

Beyond technical and fairness metrics, evaluation considers tangible administrative and clinical outcomes, such as reduced documentation time, improved diagnostic accuracy, or enhanced treatment adherence. These metrics are inherently context-dependent, making exhaustive enumeration impractical, yet they remain critical for assessing the true utility of LLMs in healthcare.

*Ultimately*, the rigorous, multi-dimensional evaluation of LLM applications in healthcare is essential for achieving safe, reasonable, and effective outcomes. By integrating quantitative and qualitative metrics, continuously monitoring biases, and engaging patients, clinicians, and other stakeholders, LLMs can align with broader healthcare goals—trustworthiness, fairness, and meaningful clinical impact. As these models evolve, flexible and context-specific evaluation frameworks will guide their responsible and sustainable integration into various healthcare ecosystems.

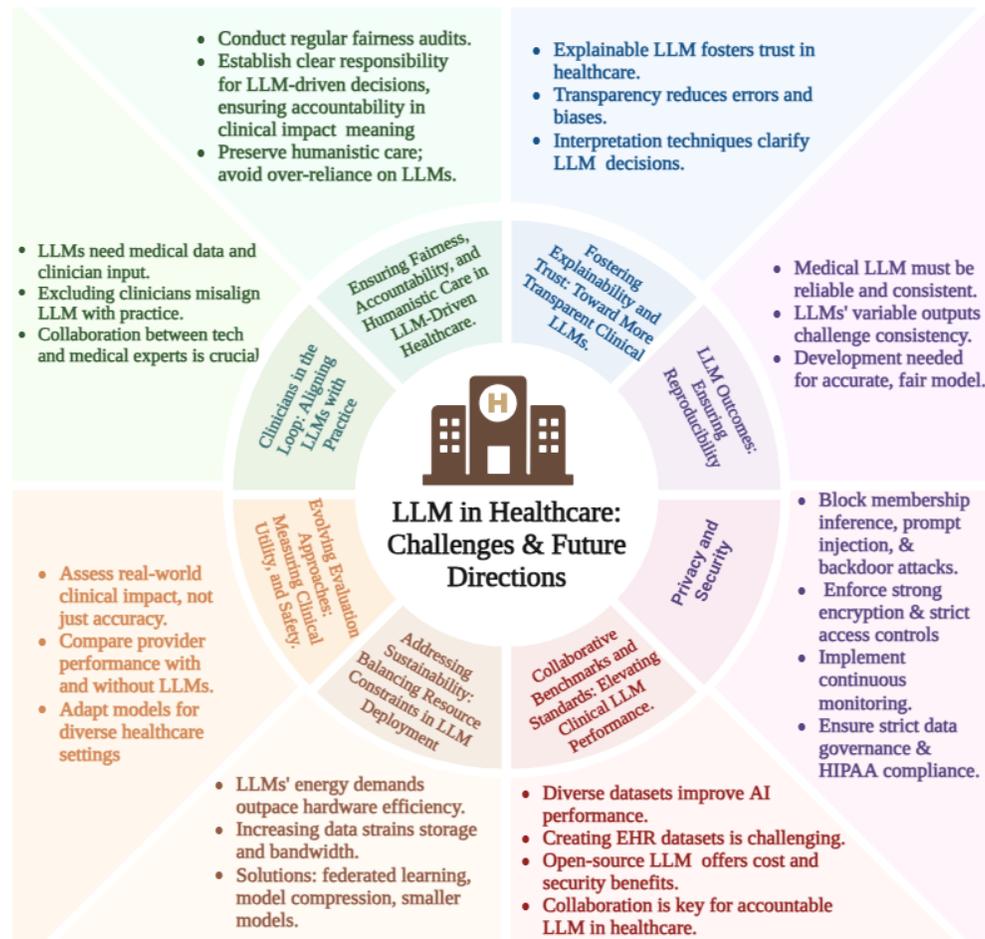

*Figure 2 : Key Challenges and Future Directions for LLMs in Healthcare*

## 4. Challenges and Opportunities for LLMs in Healthcare

LLMs have rapidly emerged as transformative tools in healthcare, demonstrating sophisticated capabilities that span clinical, operational, and educational domains. These models, built on transformer architectures and trained on extensive datasets, have shown remarkable abilities in medical text summarization, clinical documentation, decision support, and patient communication.[4,50,51] Despite these advancements, challenges remain in aligning their design and deployment with healthcare needs.[1,7] Learning from healthcare's historical challenges with IT system adoption—where insufficient clinical input led to significant workflow inefficiencies[1]—the critical importance of proactive clinician involvement in LLM development has become increasingly apparent. Studies indicate that successful implementation requires a paradigm shift: moving beyond the simple adoption of generic LLMs toward purposefully designing these tools for specific medical applications through deep integration of domain expertise. [8,52-58] This collaborative approach ensures that models truly address the nuanced demands of clinical environments. Research demonstrates that even with strong clinical partnerships, responsible LLM integration demands comprehensive consideration of multiple factors.[12] Beyond technical capabilities, success hinges on maintaining high standards of data quality, ensuring model interpretability, and developing robust infrastructure—all essential elements for creating systems that genuinely serve both patients and practitioners while seamlessly integrating into existing workflows.

The following subsections explore challenges and future directions, focusing on clinician-guided development, transparency, reliability, and ethical safeguards. Figure 2 outlines key areas for LLM integration, emphasizing ethical frameworks, clinician input, and technical reliability, while highlighting the importance of interdisciplinary collaboration to address bias, energy efficiency, and clinical validation for reasonable healthcare solutions.

**4.1 Clinicians in the Loop: Aligning LLM with Clinical Needs.**

While general-purpose LLMs demonstrate capabilities for many medically relevant tasks, their self-supervised pretraining and instruction fine-tuning overlooks crucial domain-specific data needed for optimal performance in healthcare applications. Without exposure to medical records or task-specific training, LLMs lack the precise domain adaptation required for robust medical use cases. By excluding clinicians and medical professionals from the technology design process, technology companies alone are determining the trajectory of AI's role in medicine.[1,10] Historically, healthcare has erred in letting the IT industry impose systems adoption with minimal clinical input, resulting in poor alignment between technology and practice.[1] Given how potentially disruptive LLMs may prove for activities spanning clinical documentation, decision support, operations, medical coding, and patient communication, medicine cannot afford to repeat this pattern. [1,10]

Realizing the full potential of LLMs in healthcare demands a proactive partnership between software designers and medical experts throughout training, evaluation, and implementation. [55-58] This will enable the co-development of reliable AI systems tailored to the nuanced needs of patients, providers, and clinical environments.

**4.2 Fostering Explainability and Trust: Toward More Transparent Clinical LLMs.**

Explainable AI systems are crucial in healthcare. They enable medical professionals to understand the reasoning behind AI-generated decisions, fostering trust and ensuring reliability, fairness, and alignment with patient interests.[59,60]. The rapid advancement of LLMs presents both opportunities and challenges for healthcare. These powerful models can analyze complex patterns in medical text, enabling new tools to assist physicians. While these models can analyze complex patterns in medical text to assist physicians, their statistical nature also poses risks of generating erroneous or biased outputs. Model transparency and human-centered design are essential to developing trustworthy LLM systems for medicine. Explainable models, which opened the "black box" of AI, could augment physician capabilities without replacing human judgment.

Techniques like local interpretation and retrieval augmentation offer promising avenues to clarify model reasoning in an accessible manner, ensuring outputs are understandable to both patients and healthcare providers.[61-64] However, key challenges remain in maintaining accuracy while increasing transparency and generating accessible explanations for lay users.

The responsible integration of LLMs into medicine demands a patient-centered approach. Aligning the advanced capabilities of these models with shared humanistic values is essential. To develop explainable and trustworthy LLMs in healthcare, researchers, and practitioners are recommended to focus on advancing interpretability techniques, developing robust evaluation frameworks, mitigating biases, integrating human oversight, developing interdisciplinary collaboration, establishing regulatory guidelines, and engaging in user-centered design. By prioritizing these areas, LLMs will be created to empower patients and providers, leading to more informed, personalized, and fair care delivery.[1,65]

**4.3 LLM Outcomes: Ensuring Reproducibility.**

In the healthcare domain, the reproducibility of AI systems is critical. Medical decisions often have significant consequences, and inconsistent or unpredictable AI performance could put patient safety at risk. Reproducibility is essential to ensure that LLM models generate consistent results across different settings, patient populations, and time periods. This is particularly important because if AI assists in clinical decision-making, medical professionals must be able to trust the outputs of these systems.

However, because LLMs are probabilistic algorithms, they can generate varying responses when presented with the same task or question multiple times. These responses might be improved, corrected, or simply different versions of previous responses. This poses a challenge to ensuring model reproducibility, a necessity for practical use in the medical field. Further development is needed to ensure LLMs consistently meet crucial standards of accuracy, credibility, and fairness before responsible deployment in real-world medical applications. [54,66]

**4.4 Privacy and Security.**

LLMs deployed in healthcare are uniquely vulnerable. Trained on vast datasets that often contain protected health information (PHI), they are exposed to risks like membership inference, prompt injection, and backdoor attacks. These vulnerabilities can lead to unauthorized data extraction or manipulation of clinical outputs, ultimately compromising patient safety and the reliability of decision support systems. To mitigate these risks, healthcare organizations must implement robust security measures, such as strong encryption, strict access controls, continuous monitoring, and rigorous data governance frameworks. These measures should ensure compliance with regulations like HIPAA to safeguard sensitive information and maintain system integrity.[67-69]

**4.5 Ensuring Fairness, Accountability, and Humanistic Care in LLM-Driven Healthcare.** LLMs promise to transform patient care, medical education, and clinical decision-making. Yet achieving responsible integration requires adherence to core ethical principles - *accountability, autonomy, integrity, non-maleficence, privacy, security, transparency, and trust*- to ensure safe and reasonable application.[70] <u>Accountability and Autonomy in LLM-Driven Care</u>: Accountability is paramount as LLMs enter high-stakes clinical environments, where algorithmic missteps can lead to patient harm.[67] Clinicians, developers, and institutions must share oversight to ensure that no single entity avoids responsibility for the technology's outcomes.

Simultaneously, preserving patient autonomy, with LLMs supporting informed consent by providing accessible, accurate, and unbiased information that respects patient dignity and choice.

*Fairness:* LLMs are vulnerable to amplifying biases present in their training data, potentially perpetuating or even exacerbating health disparities. [4,71,72] This risk is amplified by training data limitations, including the underrepresentation of certain populations, which can lead to variations in treatment recommendations due to differences in algorithmic performance across demographic groups. [23,73,74] Addressing these challenges requires interdisciplinary approaches to identify and mitigate biases, ensuring fair benefits across various patient populations. [14,75-77].

*Integrity and Non-Maleficence:* To ensure integrity, the use of LLMs demands rigorous validation to prevent errors or misuse. While these systems can enhance decision-making, they may generate hallucinations—fabricated but plausible outputs—that jeopardize patient safety. [67]. Ethical LLM systems must prioritize non-maleficence by deploying robust safeguards to minimize harm, including continuous monitoring of accuracy and reliability in clinical applications.

*Building Confidence in LLMs-Systems:* Building confidence in LLMs is essential for their successful integration into the clinical workflows. Models that clearly communicate their data sources, reasoning processes, and known boundaries instill a sense of trust among clinicians and patients. In contrast, black-box methods may lead to skepticism and the potential for misplaced reliance on uncertain outputs. [70,78-80] By clarifying the evidence behind decisions, healthcare systems help users understand why and how recommendations are made. *Humanistic Care:* Despite their potential efficiency, LLMs lack the emotional intelligence central to compassionate care. A humanistic approach positions these tools as supplements to, not replacements for, human judgment and empathy.[78-80] Preventing over-reliance on automation ensures that the patient-provider relationship retains its personal, empathetic core—a critical element of healing and trust.[81]

*Path Forward:* LLMs hold transformative potential for healthcare but must adhere to ethical principles to promote fairness and trust. Mitigating biases, safeguarding privacy, and preserving human judgment are essential for responsible integration, necessitating ongoing collaboration.

## 4.6 Evolving Evaluation Approaches: Measuring Clinical Utility, and Safety.

Current metrics for evaluating healthcare LLMs often focus on technical accuracy and uncertainty but these alone fall short of capturing their real-world clinical impact.[42] Demonstrating true utility necessitates rigorous studies that compare provider performance with and without these models, moving beyond technical measures to encompass patient-centered outcomes and safety. [82,83] Future evaluation frameworks must integrate assessments of clinical relevance by determining the alignment of LLM outputs with evidence-based practices; patient-centered outcomes by examining effects on health status, satisfaction, and quality of care; safety by identifying and mitigating potential risks; generalizability by systematically testing performance across varied healthcare settings; and customization through site-specific tuning and context-driven adaptations [4,7,51,83]. By adopting this comprehensive approach, we can better ensure that LLMs enhance patient care—not just by performing well in controlled settings but by effectively addressing the varied needs of real-world clinical practice.

## 4.7 Collaborative Benchmarks and Standards: Elevating Clinical LLM Performance.

*Building Robust Clinical Benchmarks*: LLMs have shown promising potential to transform healthcare by significantly improving efficiency and productivity within clinical settings. [28,84-86] However, rigorous evaluation and validation are essential before widespread deployment in real-world healthcare settings.[87] Improving LLM performance hinges on developing high-quality datasets encompassing various clinical tasks, a task necessitating extensive interdisciplinary collaboration. Building such a benchmark, particularly with EHR-based tasks and expert-validated responses, presents substantial logistical and financial hurdles. A critical need exists for an accessible EHR dataset encompassing varied queries, inpatient/outpatient settings, and both structured and unstructured data. ***Open-Source vs. Closed-Source:*** The choice between open-source and closed-source LLMs in healthcare involves critical trade-offs in cost, performance, and data security. Closed-source models often outperform open-source ones in general tasks but come with higher costs and uncertain performance in healthcare benchmarks [88,89]. Open-source LLMs can mitigate expenses and enable on-premises deployment, enhancing data security by keeping sensitive information within institutional infrastructure [88,89]. Due to their reliance on cloud processing, closed-source models necessitate stringent compliance measures such as encryption and HIPAA protocols to safeguard patient data. Future efforts should prioritize on improving LLM performance in clinical tasks, the establishment of shared training benchmarks, and stakeholder collaboration to develop open-source models that advance accountable AI in healthcare.[87,90]

**4.8 Addressing Sustainability: Balancing Resource Constraints in LLM Deployment**

The deployment of healthcare LLMs presents significant sustainability challenges due to their substantial computational demands. Evidence suggests that training and deploying models require substantial energy consumption, frequently exceeding advancements in hardware efficiency[91]. This burden is amplified in medical settings, where privacy regulations necessitate dedicated local infrastructure. The exponential growth of medical data, including imaging and electronic health records, places a significant strain on storage capacity and network resources. While technical solutions like model compression offer partial remediation, healthcare institutions require comprehensive environmental strategies that prioritize both clinical efficacy and ecological responsibility. Future developments must prioritize resource-efficient architectures optimized for medical applications.[91]

**Conclusion**

LLMs hold immense potential to revolutionize healthcare by automating clinical workflows, enhancing patient engagement, and advancing medical research and education. In this review, we have presented a framework for responsibly integrating these models into healthcare settings that emphasizes collaboration among clinicians, developers, and data scientists. Critical challenges remain around data privacy, bias mitigation, regulatory compliance, and clinical validation. By addressing these considerations through robust evaluation frameworks and interdisciplinary collaboration while maintaining focus on patient outcomes, LLMs can be responsibly leveraged to enhance healthcare delivery. Future work should focus on conducting large-scale, prospective clinical trials to validate these approaches in clinical settings and demonstrating measurable improvements in healthcare quality and accessibility.